\documentclass{article}
\usepackage{graphicx} 
\usepackage{amsmath, amssymb}
\usepackage{mathtools}
\usepackage[sort,numbers]{natbib}
\usepackage{comment}
\usepackage{amsmath,bm}
\usepackage{subfigure}
\usepackage{subcaption}
\usepackage{subfig}
\usepackage{url}
\usepackage{xcolor}
\usepackage{booktabs}
\usepackage{float}
\usepackage{caption}
\usepackage{authblk}
\title{A Novel Approach for Data Integration with Multiple Heterogeneous Data Sources}
\author[1]{Farimah Shamsi}
\author[1]{Andriy Derkach}

\date{}

\affil[1]{Department of Biostatistics and Epidemiology, Memorial Sloan Kettering Cancer Center}

\begin{document}

\maketitle
\begin{abstract}
    The integration of data from multiple sources is increasingly used to achieve larger sample sizes and enhance population diversity. Our previous work established that, under random sampling from the same underlying population, integrating large incomplete datasets with summary-level data produces unbiased parameter estimates. In this study, we develop a novel statistical framework that enables the integration of summary-level data with information from heterogeneous data sources by leveraging auxiliary information. The proposed approach estimates study-specific sampling weights using this auxiliary information and calibrates the estimating equations to obtain the full set of model parameters. We evaluate the performance of the proposed method through simulation studies under various sampling designs and illustrate its application by reanalyzing U.S. cancer registry data combined with summary-level odds ratio estimates for selected colorectal cancer (CRC) risk factors, while relaxing the random sampling assumption.
\end{abstract}
\section{ Introduction}

Statistical inference is traditionally based on detailed individual-level data collected from study participants. However, for studies with limited resources, achieving efficient and reliable estimation can be challenging. As various data sources become increasingly accessible, there is considerable interest to aggregate those sources to improve statistical efficiency even when data source does not contain measurements on the key variables or only summary-level data is available.

Addressing these challenges requires methodological developments for integrating data sources. In general two main approaches have been proposed for combining summary-level and individual-level data: one grounded in the generalized method of moments (GMM) and the other in the empirical likelihood framework \cite{imbens1994, qin2000}. \citet{chatterjee2016} developed a similar empirical likelihood approach to combine model-based summary statistics, which was later extended by \citet{zhang2020} to integrate summary data more efficiently. The data integration approaches considered in these settings typically assume that complete information on all relevant variables is available in at least one dataset. 

A more challenging scenario occurs when none of the data sources contain complete information, requiring methodological extensions to address this limitation. This setting falls within the broader framework of data fusion, which involves combining information from multiple sources that may not share common variables, with the aim of improving statistical inference and prediction.  \citet{li2020} developed a method for estimating causal effects by combining summary statistics from multiple datasets within a linear structural causal model, assuming joint normality for ordinary least squares (OLS) estimation. To mitigate bias arising in data fusion settings where no subject has complete information, \citet{evans2021} proposed a semiparametric framework of parallel inverse probability weighting (IPW) estimators that remain consistent under correctly specified outcome and data source models and encompass doubly robust estimators. Extending this framework, \citet{derkach2024integrating} introduced semi-parametric methods for mediation analysis when the exposure, mediator, and outcome are measured in separate studies, demonstrating that all parameters are identifiable and that the resulting estimators are consistent and asymptotically normal under a generalized linear model. The proposed methods generally rely on two fundamental assumptions: (1) the conditional distribution of the response variable given the predictor variables is identical across the populations providing the study and auxiliary data, and (2) the joint distribution of the predictors in these populations are also the same. These assumptions are reasonable when both the study and auxiliary information are drawn from the same population. However, as large external databases become more prevalent, assumption (1) likely remain valid, whereas assumption (2) is more likely to be violated.

In our motivating study, we aim to reassess the conclusions of \cite{derkach2024mediation} by relaxing assumptions of homogeneity between data sources (i.e. three studies are sampled from the same source population). The main question that motivating their study was whether, and to what extent, established risk factors account for the racial disparities observed in colorectal cancer (CRC) incidence across the United States. Authors proposed approach to combine  U.S. cancer registry, a U.S.-population-representative survey and summary level odds-ratio estimates, to rigorously evaluate what proportion of the difference in CRC risk between non-Hispanic Whites and Blacks is mediated by three potentially modifiable risk factors (CRC screening history, body mass index, and regular aspirin use). 

In this work, we present new framework to address potential differences in the joint distribution of the predictors between data sources. Our framework proposes to construct unbiased estimating equations by incorporating calibration weights based on the ratio of joint densities of covariates. The ratio is modeled through an exponential tilting model and its parametric form are estimated though methods of moments on the auxiliary information. Here, we present methodology that uses the reported covariance matrices of the estimates of the regression coefficients, and outline other auxilary statistics that can be used to estimate these weights.

The remainder of the paper is organized as follows. We first describe the statistical methods and discuss the theoretical properties of the resulting estimates (Section 2). We then study the estimates in finite samples using simulations (Section 3), and analyze colorectal cancer data (Section 4), Section 5 concludes the paper.

\section{ Methods}
\subsection{ Overview}
\label{overview}
We first provide notation and define the model and the parameters of interest. In this paper, we assume that we have a three sourced of information with block missingness in measurements of one or several key variables in each source. Specifically, we assume that we collected  individual-level data on three set of covariates but not outcome $Y$ with $\bm D_i = (\boldsymbol{X}_i, \boldsymbol{Z}_i,\bm C_i)$ for \(i = 1,2,...,N_{1}\) and  \(\textit{\textbf{X}} = (X_1, \ldots, X_p)'\), \(\textit{\textbf{Z}} = (Z_1, \ldots, Z_q)'\) and \(\textit{\textbf{C}} = (C_1, \ldots, C_r)'\) ,  a vector of regression estimates of association between \(\boldsymbol{X}\), \(\bm C\) and \(Y\), and  a vector of regression estimates of association between \(\boldsymbol{Z}\), \(\bm C\) and \(Y\). In the context of our motivating example, let $Y$ denote a binary indicator that is 1 if a person is diagnosed with CRC, $\bm X$ be a vector of established risk factors, $\bm Z$ be a indicator variable representing race, and $\bm C$ be a vector of common covariates.

We  assume that given these three sets of covariates $D=(\bm X,\bm Z,\bm C)$ the distribution of the outcome \(Y\) belongs to an exponential family
\begin{equation}
f (Y| \bm D; \theta, \psi) = \exp\left[\frac{Y - b(\theta)}{a(\psi)} + c (Y, \psi)\right],
\label{eq1}
\end{equation}
with
\begin{equation}
\theta = \beta_0 + {\boldsymbol{X}'\mathbf{\beta_x}} + {\boldsymbol{Z}'\mathbf{\beta_z}} + {\boldsymbol{C}'\mathbf{\beta_c}} .
\label{eq2}
\end{equation}
We let \(\boldsymbol{\beta} = (\beta_0, \mathit{\mathbf{\beta_x}},\mathit{\mathbf{\beta_z}},\mathit{\mathbf{\beta_c}})'\) denote all the regression parameters in (\ref{eq2}) that we are interested to estimate. 

In addition to individual level measurements on the covariates $\bm D$ from the the first study, we are provided the estimates of the associations between $\boldsymbol{X}$, $\boldsymbol{C}$ and $Y$ from the second study.  In other words, the second study collected data on \((Y, \boldsymbol{X},\bm C)\), and assumed following working model
\begin{equation}
    f(Y|\boldsymbol{X},\bm C; \theta_X, \psi_X) = \exp\left[\frac{Y\theta_X - b(\theta_X)}{a(\psi_X)} + c(Y, \psi_X)\right]
    \label{eq_working_m1}
\end{equation}
with \(\theta_X = \gamma_{0} + \boldsymbol{X}'\gamma_X+ \boldsymbol{C}'\gamma_C\) and provided the vector of estimates $\hat{\bm\gamma}=(\hat{\gamma_{0}},\hat{\bm \gamma}_{X},\hat{\bm \gamma}_{C})'$. We also assume that third study also collected data on \((Y, \boldsymbol{Z},\bm C)\), and provided estimates $\hat{\bm\alpha}=(\hat{\alpha_{0}},\hat{\bm \alpha}_{Z},\hat{\bm \alpha}_{C})'$ under following working model
\begin{equation}
    f(Y|\boldsymbol{Z},\bm C; \theta_Z, \psi_Z) = \exp\left[\frac{Y\theta_Z - b(\theta_Z)}{a(\psi_Z)} + c(Y, \psi_Z)\right]
      \label{eq_working_m2}
\end{equation}
with \(\theta_Z = \alpha_{0} + \boldsymbol{Z}'\alpha_Z+ \boldsymbol{C}'\alpha_C\).

The last assumption that we make in this section is related how these estimates were produced. By denoting the gradient operator by \(\nabla\), which yields the vector of partial derivatives, the estimate from these working models were obtained using corresponding score vectors
\begin{equation}
   U_2 (Y|\boldsymbol{X},\bm C;\boldsymbol{\gamma}) = \nabla_{\boldsymbol{\gamma}} log {f(Y|\boldsymbol{X},\bm C,\theta_X, \psi_X)} = \left[Y-b'(\theta_X)\right][1,{\boldsymbol{X}},{\boldsymbol{C}}]',
   \label{u2}
\end{equation}
and
\begin{equation}
   U_3 (Y|\boldsymbol{Z},\bm C;\boldsymbol{\alpha}) = \nabla_{\boldsymbol{\alpha}} log {f(Y|\boldsymbol{Z},\bm C,\theta_Z, \psi_Z)} = \left[Y-b'(\theta_Z)\right][1,{\boldsymbol{Z}},{\boldsymbol{C}}]',
   \label{u3}
\end{equation}
with \(\theta_X = \gamma_{0} + \boldsymbol{X}'\gamma_X+ \boldsymbol{C}'\gamma_C\) and \(\theta_Z = \alpha_{0} + \boldsymbol{Z}'\alpha_Z+ \boldsymbol{C}'\alpha_C\).

In the context of our motivating CRC example, $Y$ denotes a binary indicator that is 1 if a person is diagnosed with CRC, $\bm X$ is a vector of established risk factors, $\bm Z$ is a indicator variable representing race, and $\bm C$ is a vector of common covariates. The individual level measurements on covariates $\bm D$ were obtained from rom the U.S. National Health Interview Survey (NHIS), the second study that estimates of sex- and age-specific CRC incidence rates for different races was the National Cancer Institute’s (NCI’s) Surveillance, Epidemiology, and End Results (SEER) cancer registry database, and the third study that evaluated established risk factors  was a population-based age-matched case-control study\cite{freedman2009colorectal}. 

\subsection{Estimation based on between study homogeneity assumptions}
\label{homosection}

 To estimate $\boldsymbol{\beta}$ from $\hat{\bm\gamma}$, $\hat{\bm \alpha}$ and a dateset of individual-level measurements $\bm D$, we  assumed that all three sources of information are based on samples from the same underlying population \cite{derkach2024,derkach2024mediation}. Then we relied on the observation expectations of the score vectors (\ref{u2}) and (\ref{u3}) can be used to convert the external marginal estimate to the system of equations \cite{white82,chatterjee2016},
\begin{equation*}
    E\left[ U_2 (Y|\boldsymbol{X},\boldsymbol{C};\boldsymbol{\gamma})\right] =\int \left[b'(\theta)-b'(\theta_X)\right][1,{\boldsymbol{X}},\boldsymbol{C}]'f_2(\bm D)d\bm D=0,
\end{equation*}
and
\begin{equation*}
    E\left[ U_3 (Y|\boldsymbol{Z},\boldsymbol{C};\boldsymbol{\alpha})\right] =\int \left[b'(\theta)-b'(\theta_Z)\right][1,{\boldsymbol{Z}},\boldsymbol{C}]'f_3(\bm D)d\bm D=0,
\end{equation*}
where $\theta = \beta_0 + {\boldsymbol{X}'\mathbf{\beta_x}} + {\boldsymbol{Z}'\mathbf{\beta_z}} + {\boldsymbol{C}'\mathbf{\beta_c}}$; 
$f_2(\bm D)$ and $f_3(\bm D)$ are study specific joint distributions of $\bm D$. 

In the setting we considered there, the joint distribution of covariates is not presented to us from the study 2 and 3. Under assumption that studies are random sample from the same underlying population, i.e. empirical distribution of $\bm D$ is the same across three studies, \citet{derkach2024} proposed to "plug in" empirical distribution $\hat{F}_1$ from the the first study into above systems to get estimate of $\mathbf{\beta}$. As result, vector of estimates, $\hat{\bm\beta}$ is obtained by solving a combination of system of equations 
\begin{equation*}
    \sum_{i=1}^{N_1} \left[b'(\hat{\theta}_i )-b'(\hat{\theta}_{X,i})\right][1,{\boldsymbol{X}}_i,{\boldsymbol{C}}_i]'=0,
\end{equation*}
and
\begin{equation*}
    \sum_{i=1}^{N_1} \left[b'(\hat{\theta}_i )-b'(\hat{\theta}_{Z,i}))\right][1,{\boldsymbol{Z}}_i,{\boldsymbol{C}}_i]'=0,
\end{equation*}
where $\hat{\theta}_i =\hat{\beta}_0 + {\boldsymbol{X}'}_i\mathbf{\hat{\beta}_x} + {\boldsymbol{Z}'}_i\mathbf{\hat{\beta}_z}+ {\boldsymbol{C}'}_i\mathbf{\hat{\beta}_c} $,~\(\hat{\theta}_{X,i} = \hat{\gamma}_{0} + \boldsymbol{X}_i'\hat{\gamma}_X+\bm C_i\hat{\gamma}_C\), and  \(\hat{\theta}_{Z,i} = \hat{\alpha}_{0} + \boldsymbol{Z}_i'\hat{\alpha}_Z+\bm C'_i\hat{\alpha}_C\)
However, if the assumption of random sampling is violated, the proposed approach may yield biased estimates \cite{chatterjee2016,derkach2024,kundu2019generalized}; even when the conditional distribution  $f(Y|\bm D)$ is the same across three studies and only joint distribution of covariates is different $f_2(\bm D)\neq f_1(\bm D)$ or $f_3(\bm D)\neq f_1(\bm D)$.

\subsection{Estimation based on between study heterogeneity assumptions}
\label{hetsection}
In this section, we introduce our framework to integrate heterogeneous data sources. We assume that $f (Y| \bm D; \theta, \psi)$, conditional distribution of $Y$ given $\bm D$, is the same across three source, and only joint distribution of covariates  $f(\bm D)$ may differ across three studies. These assumption correspond to setting called covariate shift in the machine learning literature\cite{tibshirani2019conformal,lu2024combining}.

Under these assumptions, the calibration weights based on density ratios  given by
\begin{equation}
    w_2 (\bm D) = \frac{f_2(\bm D)}{f_1(\bm D)} \text{ and }     w_3 (\bm D) = \frac{f_3(\bm D)}{f_1(\bm D)},
\label{w3}
\end{equation}
transform equations (\ref{u2}) and (\ref{u3}) to the unbiased equations while integrating over the distribution of the data from the study 1 
\begin{equation*}
    \int \left[b'(\theta)-b'(\theta_X)\right][1,{\boldsymbol{X}},\bm C]'w_2 (\bm D)f_1(\bm D)d\boldsymbol{D}=0,
\end{equation*}
and
\begin{equation*}
   \int\left[b'(\theta)-b'(\theta_Z)\right][1,{\boldsymbol{Z}},\bm C]'w_3 (\bm D)f_1(\bm D)d\bm D=0,
\end{equation*}
where $\theta$, $\theta_X$ and $\theta_Z$ are defined in (\ref{eq2}-\ref{eq_working_m2}). In this section, we propose new approaches to model these ratios under setting studied here.

Let \(R_j \in \{0,1\}\) denote the observation indicator such that \(R_j = 1\) if a subject is  sampled into a particular study $j \in \{2,3\}$ and \(R_j = 0\) otherwise. Let \(\pi_2 (\bm D) = Pr(R_2=1|\bm D)\) be the probability that a subject is in study 2, \(\pi_3 (\bm D) = Pr(R_3=1|\bm D)\) be the probability that a subject is sampled to study 3. In addition, without loss of generality, we assume that study one is a source population from which observations are sampled to the other two studies with \(0<\pi_2(\bm D)<1\) and \(0<\pi_3(\bm D)<1\). The set of these assumptions are closely related to assumptions of consistency, transportability, and positivity made in causal literature \cite{vo2019novel}. 

To model the density ratios in (\ref{w3}), we assume that we can correctly specify them a parametrically through exponential tilting models in sense that   
\begin{equation}
    w_j (\bm D)=\frac{f_j(\bm D)}{f_1(\bm D)}=\frac{\pi_j (\bm D)}{P(R_j=1)} =\exp(\phi(\bm D;\zeta_j)) \text{ for } j\in\{2,3\}
    \label{exp2}
\end{equation}
where $\phi(\bm D;\zeta_j)$ are specified parametric models with unknown parameters $\zeta_j$ for $j \in \{2,3\}$. In the presence of observed covariate data from all three studies, $phi(\bm D;\zeta)$ can be modeled flexibly through logistic regression approaches with linear combinations of predictors, polynomial terms, or splines. In the next sections, we present novel approaches to estimate the specified parametric models $\phi(\bm D;\zeta_2)$ and $\phi(\bm D;\zeta_3)$ using auxiliary data that is usually provided in addition to the vector of the estimates $\hat{\bm \gamma}$ and $\hat{\bm \alpha}$.

\subsubsection{Incorporation of the covariance matrices of $\hat{\bm \gamma}$ and $\hat{\bm \alpha}$ }
\label{Cov_section}
In this section, we describe methods that incorporate auxiliary information to estimate adjustments for between source heterogeneity using the framework described above. In particular, we propose to utilize covariance matrices of the estimates from the studies 2 and 3, $\hat{Cov}(\hat{\bm \gamma})$ and $\hat{Cov}(\hat{\bm \alpha})$, as the auxiliary information sources. In the Section \ref{sec:other_aux}, we briefly describe additional approach that uses regression estimates from the set of univariate regressions that are commonly reported as initial analysis in the main paper or supplement.

Under working models (\ref{eq_working_m1}) and (\ref{eq_working_m2}), covariances of the reported estimates $\hat{\bm \gamma}$ and $\hat{\bm \alpha}$ are proportional to the inverse of observed Fisher information 
\begin{equation*}
  \hat{Cov}(\hat{\bm \gamma})^{-1}  = -\frac{1}{a(\hat{\psi}_X)}\sum_{i=1}^{N_2} b''(\hat{\theta}_{X,i})[1,{\boldsymbol{X}}_i,\bm C_i]'[1,{\boldsymbol{X}}_i,\bm C_i] = N_2 \hat{I}_2(\hat{\theta}_X,\hat{\psi}_X),
    \label{OF_gamma}
\end{equation*}
and 
\begin{equation*}
   \hat{Cov}(\hat{\bm \alpha})^{-1}  = -\frac{1}{a(\hat{\psi}_Z)}\sum_{i=1}^{N_3} b''(\hat{\theta}_{Z,i})[1,{\boldsymbol{Z}}_i,\bm C_i]'[1,{\boldsymbol{Z}}_i,\bm C_i] =N_3 \hat{I}_3(\hat{\theta}_Z,\hat{\psi}_Z),
   \label{OF_alpha}
\end{equation*}
with $\theta_{X,i}$ and $\theta_{Z,i}$ are defined above. Note that to these covariance matrices are calculated using reported estimates of regression coefficients and observed individual level data in the studies 2 and 3 that are not given to us. 

We propose to plug in individual level data from the study 1 under exponential tilting models (\ref{exp2}) into above equations to re-estimate observed Fisher's information
\begin{equation*}
   \hat{J}(\hat{\theta}_X,\hat{\psi}_X;\zeta_2) = -\frac{1}{a(\hat{\psi}_X)}\sum_{i=1}^{N_1}  \exp\left[\phi(\bm D_i;\zeta_2)\right]  b''(\hat{\theta}_{X,i})[1,{\boldsymbol{X}}_i,\bm C_i]'[1,{\boldsymbol{X}}_i,\bm C_i],
    \label{OF_gamma_st1}
\end{equation*}
and 
\begin{equation*}
  \hat{J}(\hat{\theta}_Z,\hat{\psi}_Z;\zeta_3) =  -\frac{1}{a(\hat{\psi}_Z)}\sum_{i=1}^{N_1}  \exp\left[\phi(\bm D_i;\zeta_3)\right] b''(\hat{\theta}_{Z,i})[1,{\boldsymbol{Z}}_i,\bm C_i]'[1,{\boldsymbol{Z}}_i,\bm C_i],
   \label{OF_alpha_st1}
\end{equation*}
and compare (after appropriate sample size scaling) these matrices with reported $\hat{I}_2(\hat{\theta}_X,\hat{\psi}_X)$ and $\hat{I}_3(\hat{\theta}_Z,\hat{\psi}_Z)$. If the exponential tilting models transforms distribution of covariates to those in the studies in 2 and 3, the distance between these matrices will be small and large otherwise.

As result, to estimate the vectors $\zeta_2$ and $\zeta_3$, we formulate an optimization problem that minimizes the distances between observed information matrices and estimated one (i.e. moment match). Specifically, the estimators are obtained as
\begin{equation}
    \hat{\zeta}_{2} = \underset{\zeta_2}{\mathrm{argmin}} \parallel \hat{I}_2(\hat{\theta}_X,\hat{\psi}_X) - 1/N_1\hat{J}(\hat{\theta}_X,\hat{\psi}_X;\zeta_2)\parallel _F^{2},
      \label{sums_sq2}
\end{equation}
\begin{equation}
    \hat{\zeta}_{3}  = \underset{\zeta_3}{\mathrm{argmin}} \parallel \hat{I}_3(\hat{\theta}_Z,\hat{\psi}_Z) - 1/N_1\hat{J}(\hat{\theta}_Z,\hat{\psi}_Z;\zeta_3)\parallel _F^{2},
    \label{sums_sq3}
\end{equation}
where \(\parallel . \parallel_F\) denotes the Frobenius norm, defined as the square root of the sum of squared matrix elements.  

\textit{Remark 1}: Above optimization problems can be viewed in the context of regression, where elements of the Fisher information matrices $\hat{I}_2(\hat{\theta}_X,\hat{\psi}_X)$ and $\hat{I}_3(\hat{\theta}_Z,\hat{\psi}_Z)$ are observed outcomes and elements of 
$\hat{J}(\hat{\theta}_X,\hat{\psi}_X;\zeta_2)$  and $\hat{J}(\hat{\theta}_Z,\hat{\psi}_Z;\zeta_3)$ are fitted values. Thus, there is restriction on the complexities of the parameterization of tilting models represented by parameters $\zeta_2$ and $\zeta_3$. In general, since, the observed Fisher information matrices are symmetric, the maximum size of the vector $\zeta_2$ is $l(l+1)/2$ and maximum size of the vector $\zeta_3$ is $m(m+1)/2$ (i.e. number of unique values in symmetric matrix  $\hat{I}_2(\hat{\theta}_X,\hat{\psi}_X)$ and $\hat{I}_3(\hat{\theta}_Z,\hat{\psi}_Z)$). 

\textit{Remark 2}: Framing of estimation of $\zeta_2$ and $\zeta_3$ as regression tasks enables the use of a train–test framework for selecting the optimal weighting model from several candidates functions $\phi(\bm D;\zeta_2)$. Proposed  method involves splitting the $l(l+1)/2$ or $m(m+1)/2$ elements of Fisher's information matrices (e.g. 80/20 ratio) into training and testing subsets. During the training step, candidate models are fit on  entries of $\hat{I}_2(\hat{\theta}_X,\hat{\psi}_Z)$  and $\hat{I}_3(\hat{\theta}_Z,\hat{\psi}_Z)$ selected for training. During the testing step, the fitted models are evaluated by measuring distance between observed and predicted values on the testing entries of information matrices. In final step, the selected weighting model is applied to entire Fisher information matrix to get final estimate of ${\bm \zeta}_2$ and ${\bm \zeta}_3$.

\textit{Remark 3}: Typically, the dispersion factors such as $\hat{\psi}_X$ and $\hat{\psi}_Z$ are not reported in practice. We propose to set them to $1$ as they are scalars and are absorbed into the intercept of weighting function

 After $\hat{\zeta}_2$ and $\hat{\zeta}_3$ are obtained by above approaches, the estimate $\hat{\bm\beta}$ is obtained by solving a combination of weighted system of equations 
\begin{equation}
    \sum_{i=1}^{N_1} \exp\left[\phi(\bm D_i;\hat{\zeta}_2)\right]\left[b'(\hat{\theta}_i)-b'(\hat{\theta}_{X,i})\right][1,{\boldsymbol{X}}_i,\bm C_i]'=0,
    \label{estim_w1}
\end{equation}
and
\begin{equation}
\sum_{i=1}^{N_1}\exp\left[\phi(\bm D_i;\hat{\zeta}_3)\right]\left[b'(\hat{\theta}_i)-b'(\hat{\theta}_{Z,i})\right][1,{\boldsymbol{Z}}_i,\bm C_i]'=0.
      \label{estim_w2}
\end{equation}

\subsection{Incorporation of the covariance matrices under case-control sampling}

In this section, we provide extension of our work when summary level data is provided from the case-control study. Here, we assume without loss of generality that, study 2 is a case-control study. Let $N^1_2$ and $N^0_2$ represent a known number of cases and controls. The reported covariance matrix from the logistic model (\ref{eq_working_m1}), $Cov(\hat{\alpha})$, has slightly different form then under prospective sampling,
\begin{eqnarray*}
  \hat{Cov}(\hat{\bm \gamma})^{-1}  &=& \sum_{i=1}^{N^1_2} \frac{\exp(\hat{\theta}_{X,i})}{(1+\exp\hat{\theta}_{X,i}))^2}[1,{\boldsymbol{X}}_i,\bm C_i]'[1,{\boldsymbol{X}}_i,\bm C_i]\\ &+& \sum_{i=1}^{N^0_2} \frac{\exp(\hat{\theta}_{X,i})}{(1+\exp(\hat{\theta}_{X,i}))^2}[1,{\boldsymbol{X}}_i,\bm C_i]'[1,{\boldsymbol{X}}_i,\bm C_i] = (N^1_2+N^0_2) \hat{I}^R_2(\hat{\theta}_X,\hat{\psi}_X),
\end{eqnarray*} 

where \(\hat{\theta}_{X,i} = \hat{\gamma}^*_{0} + \boldsymbol{X}_i'\hat{\gamma}_X+\bm C_i\hat{\gamma}_C\) and $\gamma^*_0$ is an intercept adjusted for the case-control selection.

Under the case-control sampling, patients are sample to the study based observed outcomes $Y$ from underlying study population. Thus, the distribution of the vector the vector of covariates $f_2(\bm D)$ in study 2 is a mixture of two distributions $f(\bm D|Y=1,R_2=1)$ in cases and $f(\bm D|Y=0,R_2=1)$ in controls. These distributions can be further rewritten using Bayes' rule
\begin{equation*}
    f(\bm D|Y,R_2=1) = \frac{f(Y|\bm D;\theta,\psi)\pi_2(\bm D)f(\bm D)}{f(Y|R_2=1)P(R_2=1)}=\exp(\phi(\bm D;\zeta_2))\frac{f(Y|\bm D;\theta,\psi)f(\bm D)}{f(Y|R_2=1)},
\end{equation*}
where $f(Y|\bm D;\theta,\psi)$ is logistic model in (\ref{eq1}). To simplify further, we assume that outcome $Y$  is rare in population and as result $f(Y=0|\bm D;\theta,\psi)\approx 1$ and $f(Y=1|\bm D;\theta,\psi)\approx \exp(\theta)$. This assumption allow us to rewrite $f(\bm D|Y=0,R_2=1)\approx f(\bm D|R_2=1)=\exp(\phi(\bm D;\zeta_2))f(\bm D)$ and 
\begin{equation*}
    f(\bm D|Y=1,R_2=1)=\frac{\exp(\phi(\bm D;\zeta_2))}{P(Y|R_2=1)}\exp(\theta)f(\bm D).
\end{equation*}
As result distribution of covariates in cases is subject to two transformations: one due to selection into study 2 and the second due to selection based on outcome. The estimates Fisher information matrix based on study 1 data can  written as in the following form
\begin{eqnarray*}
   \hat{J}^R(\hat{\theta}_X,\hat{\psi}_X;\zeta_2,\bm \beta) &=& \rho\sum_{i=1}^{N_1}  \frac{\exp(\phi(\bm D;\zeta_2))}{P(Y|R_2=1)}\exp(\phi(\bm D_i;\zeta_2))\frac{\exp(\hat{\theta}_{X,i})}{(1+\exp(\hat{\theta}_{X,i}))^2}[1,{\boldsymbol{X}}_i,\bm C_i]'[1,{\boldsymbol{X}}_i,\bm C_i]\\ &+& (1-\rho)\sum_{i=1}^{N_1}  \exp(\phi(\bm D_i;\zeta_2))\frac{\exp(\hat{\theta}_{X,i})}{(1+\exp(\hat{\theta}_{X,i}))^2}[1,{\boldsymbol{X}}_i,\bm C_i]'[1,{\boldsymbol{X}}_i,\bm C_i].
\end{eqnarray*}
where $\rho = N^1_2/(N^1_2 + N^0_2)$. As result $\hat{J}^R(\hat{\theta}_X,\hat{\psi}_X;\zeta_2,\bm \beta)$ is function of both $\bm \beta$ and $\bm \zeta_2$ that has to be estimated iteratively. 

Specifically, we propose to estimate $\zeta_2$ based on the preliminary value of $\hat{\beta}^{k-1}$ (i.e. starting value of $\hat{\bm \beta}^0$ is set to the estimate without calibration weights)
\begin{equation}
    \hat{\zeta}^{k}_{2} = \underset{\zeta_2}{\mathrm{argmin}} \parallel \hat{I}^R_2(\hat{\theta}_X,\hat{\psi}_X) - 1/N_1\hat{J}^R(\hat{\theta}_X,\hat{\psi}_X;\zeta_2,\hat{\beta})\parallel _F^{2},
\end{equation}
then estimate $\hat{\beta}^{k}$ is obtained using system of equations (\ref{estim_w1}) and (\ref{estim_w2}) with $\hat{\zeta}^{k}_{2}$ and $\hat{\zeta}_{3}$ plugged to weighting functions. This iterative process is continued until the convergence is reached. 

\textit{Remark 3}: Given that true intercept of (\ref{eq_working_m1}) is not estimable under the case-control setting, equation (\ref{estim_w1}) has to be rewritten to account it
\begin{equation*}
    \sum_{i=1}^{N_1} \exp\left[\phi(\bm D_i;\hat{\zeta}_2)\right]\left[b'(\hat{\theta}_i)-b'({\gamma^{**}_0} + {\boldsymbol{X}'_i\mathbf{\hat{\gamma}_X}}+\bm C'_i\hat{\bm \gamma}_C)\right][1,{\boldsymbol{X}}_i,\bm C_i]'=0,
\end{equation*}
where $\gamma^{**}_0$ is additional nuisance parameters \cite{derkach2024integrating,derkach2024} that will be estimated along with vector $\bm \beta$.
\subsection{Incorporation of other types of auxiliary information}
\label{sec:other_aux}
In this section, we briefly outline approach that estimate $\bm \zeta_2$ and $\bm \zeta_3$ using the vector of estimates from univariable models, that are usually reported in the main paper alongside multivariable estimates. Let $(\hat{\alpha}^M_{0,k},\hat{\alpha}^M_{X,k})$ be the reported estimate of association between $Y$ and a covariate $X_k$ for $k=1,...,p$. To determine calibration weight, we use the fact that the marginal and full model estimates $\hat{\alpha}^M_{X,k}$ and $\hat{\bm \alpha}$ are connected through the fact that 
$$
  \int \left[b'(\theta^M_k)-b'(\theta_X)\right][1,X_k]'f_2(\bm D)d\bm D=0,
$$ where $\theta^M_k = \alpha^M_0 + \alpha^M_{X,k}X_k$ and $\theta_M$ is defined in (\ref{eq_working_m1}). Given this observation we can construct estimation equations to determine optimal value of $\bm \zeta_2$
$$
  \int \exp\left[\phi(\bm D_i;{\zeta}_2)\right]\left[b'(\theta^M_k)-b'(\theta_X)\right][1,X_k]'f_1(\bm D)d\bm D=0 \text{ for } k=1,..,p
$$
\subsection{Statistical inference}
In this section, we outline asymptotic covariance matrix of the vector of estimates $\hat{\bm \beta}$ obtained using estimating equations  (\ref{estim_w1}) and (\ref{estim_w2}). In practice with several candidate weighting functions $\phi(\bm D;\zeta)$, we propose to use parametric bootsrap. At each bootstrap replication, we resample individual-level data $(\bm X, \bm Z, \bm C)$ with replacement, and simulate new values of $\hat{\bm \alpha}$ and $\hat{\bm \gamma}$ from the asymptotic distributions $N(\bm \alpha,Cov({\bm \alpha}))$ and $N(\bm \gamma,Cov({\bm \gamma}))$. Then we estimate $\hat{\bm \beta}^b$ using generated data and our proposed approach with selecting best weighting function.

In the case of single candidate model $\phi(\bm D;\zeta)$, to calculate covariance of the vector of estimates $\hat{\bm \eta}=(\hat{\beta},\hat{\zeta}_2,\hat{\zeta}_3)'$ , we will rely  on theory of estimating equations. Let $\bm S_\beta(\bm D,\bm\eta)$ be set of estimating  equations for $\bm \beta$ based on the linear combination of equations in (\ref{estim_w1}) and (\ref{estim_w2}). Let $\bm S_{\zeta_2}(\bm D,\bm \eta)$ and $\bm S_{\zeta_3}(\bm D,\bm \eta)$ be vectors of estimating equations for the  $\bm \zeta_2$ and $\bm \zeta_2$, that are derived by taking partial derivatives of (\ref{sums_sq2}) and (\ref{sums_sq3}). Based on the  Theorem 1 of Yuan and Jennrich (2000), asymptotic distribution of $\hat{\bm \eta}$ can be summarized by following proposition

\textit{Proposition 1.} Assume that   $N_1 \to \infty $ then under the model in (\ref{eq1}) and (\ref{eq2}), the solution  $\hat{\bm \eta}$ of the equations (\ref{sums_sq2}-\ref{estim_w2}) satisfies 
\begin{equation}
\label{eq:prop2}
\lim\limits_{N_1 \to \infty} \sqrt{N_1}\left(\hat{\bm \eta}-  \bm \eta\right) \xrightarrow[d]{} N\left(\bm 0,J^{-1} B(\bm \eta) J^{-1} \right),
\end{equation}
where $\bm S(\bm D,\bm \eta)=(\bm S_\beta(\bm D,\bm \eta),\bm S_{\zeta_2}(\bm D,\bm \eta),\bm S_{\zeta_3}(\bm D,\bm \eta)$), $B(\bm \eta) = E( \bm S(\bm D,\bm \eta)\bm S(\bm D,\bm \eta)')$ and $J = E ( \nabla_{\bm \eta}  \bm S(\bm D,\bm \eta))$.

Under this derivations, we assumed that $\hat{\bm \alpha}$ and $\hat{\bm \beta}$ are estimated with sufficiently large sample sizes that their variability is negligible. Extension to incorporate uncertainly in these vectors is straight forward.

\section{Simulation studies}

We evaluate our proposed procedures for estimation of regression coefficients in multiple simulation studies. In these studies, we compare the performance of three methods for data integrations under two data generating models and three bias sampling scenarios. In each study, we assume that we have three sources of information: 1) a dataset containing individual level data $\bm D_i$ for $i=1,...,N$, 2) a vector of estimates from the regression of  $\bm X$ and $\bm C$ on $Y$ obtained from the study of the size $N$, and 3)  a vector of estimates from the regression of $\bm Z$ and $\bm C$ on $Y$ obtained from the study of the same size $N$. In the main paper, we set $N=1000$ and $N=10000$ in supplement materials.

\subsection{Underlying data generating model}
\label{udgmodel}
We consider a binary outcome $Y$ and continuous set of eight covariates $\bm D$ that is composed of three vectors $\bm X=(X_1,X_2,X_3)$, $\bm Z=(Z_1,Z_2,Z_3)$ and $\bm C=(C_1,C_2)$. We assume that the underlying model for the outcome is given by
\[
logit \{P(Y_i|\bm D)\}= \mathbf{\beta_0} + {\boldsymbol{X}'}_i\mathbf{{\beta}_x}+ {\boldsymbol{Z}'}_i\mathbf{{\beta}_z} + {\boldsymbol{C}'}_i\mathbf{{\beta}_c},
\]
where $\mathbf{\beta_0}=-3$, $\mathbf{{\beta}_x} = (0.5, 0.25, -0.5) $, $\mathbf{{\beta}_z} = (0.25, 0.5, -0.25)$, $\mathbf{{\beta}_c} = (0.25, 0.25)$.
 
 We evaluate our approaches under two joint distribution of the covariates. Under \textit{the first data generating model}, we assume that the  vector $ \boldsymbol{D}$ follows a multivariate normal distribution, 
$
D_i \sim  N_{8} (\boldsymbol{\mu}, \boldsymbol{\Sigma})
$ with mean vector $\boldsymbol{\mu}=(0.25,0.5,0.25,0.5,0.25,0.5,0.25,0.5)$ and the covariance matrix $\boldsymbol{\Sigma}$ is such that diagonal elements are set to 1 and  off-diagonal elements are set to 0.5. Under this setting, all covariates have pairwise linear relationship. Under \textit{the second data generating model}, we first generate $\boldsymbol{D}^* = (\bm X^*,\bm Z,\bm C)$ from the same multivariate normal distribution and then set $\bm X$ to $|\bm X^*|$, hence, providing non-linear relationship between $\bm Z$, $\bm C$ and $\bm X$. 

\subsection{Bias sampling scenarios}
 
 We also consider three scenarios of sampling  of the data for the studies two and three are generated. In all these three scenarios, we assume  that for the first study,  covariate data $\bm D$ was generated from the model above. For the the second and the third studies, we first generate large cohort using underlying generating model presented in above Section \ref{udgmodel}, then we sample $N$ observations to each study with selection probabilities  \begin{equation*}
    \pi_2(\bm D) = \exp(\delta_2)/(1+\exp(\delta_2)) \text{ and } \pi_3(\bm D) = \exp(\delta_3)/(1+\exp(\delta_3)),
\end{equation*}
where $\delta_2$ and $\delta_3$ depends on $\bm D$. We investigate three scenarios for the forms of $\delta_2$ and $\delta_3$

\textbf{Scenario 1 : Additive model.} We assume that sampling probabilities to the studies two and three depend only on the same set of covariates as regression models through linear function . Specifically we assume $\delta_2 = \zeta_0+\bm X'\zeta_X+\bm C'\zeta_C$ and  $\delta_3 = \zeta_0+\bm Z'\zeta_Z+\bm C'\zeta_C$ with $\zeta_0 = -6$, $\bm \zeta_X=\bm \zeta_Z=(0.5, 0.5, 0.5)$ and $\zeta_C=(0.25, 0.25)$. 

\textbf{Scenario 2 : Additive model with interaction.} We also assume that sampling probabilities to the studies two and three depend only on the same set of covariates as regression models through linear function incorporating also interactions between $\bm X$ and $\bm C$ and $\bm Z$ and $\bm C$. Specifically we assume $\delta_2 = \zeta_0+\bm X'\zeta_X+\bm C'\zeta_C + \zeta_I\sum_{l>m} X_lX_m+\zeta_I\sum_{l,m} X_lC_m$ and  $\delta_3 = \zeta_0+\bm Z'\zeta_Z+\bm C'\zeta_C++ \zeta_I\sum_{l>m} Z_lZ_m+\zeta_I\sum_{l,m} Z_lC_m$. We set all interactions to the same value $\zeta_I = (0.25, 0.25, 0.25) $. 

\textbf{Scenario 3 : Full model} Lastly, we assume that sampling probabilities to the studies two and three depend only on the full set of covariates through linear function. Specifically we assume $\delta_2 =\delta_3= \zeta_0+\bm X'\zeta_X+\bm Z'\zeta_Z+\bm C'\zeta_C$,  with $\zeta_0 = -6$, $\bm \zeta_X=(0.5, 0.5, 0.5)$, $\zeta_C=(0.25, 0.25)$, and $\bm \zeta_Z =(0.25, 0.25, 0.25)$.

\subsection{Data generation in case-control studies}
 We also consider a setting in which the summary-level data for the study 2 is derived from a case–control study. Data generation procedures for other studies follows the same steps as above. To generate summary level data for case-control study, we first generate large cohort using underlying generating model presented in above Section \ref{udgmodel}, then we sample observations to initial population with selection probability  $ \pi_2(\bm D) = \exp(\delta_2)/(1+\exp(\delta_2)).$ In step 2, we sample $N/2$ cases and controls from the new population.

\subsection{Methods under investigation}
To assess the ability of our proposed method to address heterogeneity across data sources, we compare its performance with other two approaches. The first method, that we compare to, uses original method under assumption of homogeneity (see Section \ref{homosection}). The second method corresponds to the true weighting approach, in which the sampling weights are set $\exp(\delta_2)$ or$\exp(\delta_3)$. This method serves as the benchmark for evaluating the performance of the proposed weighting approaches. Last method that we evaluate in the simulation is our new approach. Here, we assume that there are two candidates  for the weighting function. The first function assumes that selection probability depends only though additive model, and the second function assumes that selection probability depends only though additive model with interaction. Lastly, the best weighting function is obtained using a train–test framework outlined in Remark 2 of Section \ref{Cov_section}. Specifically, we use 80\% of Fisher's information matrix to train both weighting models, and then select the best model using remaining  20\% of Fisher's information matrix.    

\subsection{Results}
We evaluate the performance of data integration methods by summarizing simulation results for the estimates of regression coefficients (with the intercept excluded). Figure 1 shows the empirical distribution of $(\hat{\boldsymbol{\beta}}_X, \hat{\boldsymbol{\beta}}_Z, \hat{\boldsymbol{\beta}}_C)$ obtained from three data integration methods under additive bias sampling scenario and linear relationship between covariates (panel a) and non-linear relationship between covariates (panel b). Results under the homogeneity assumption are shown in red, the benchmark using known sampling weights in green, and our proposed weighting method in blue. Under a linear relationship between $\boldsymbol{X}$ and $\boldsymbol{Z}$, both the our proposed method and the method under homogeneity assumption yield similar, unbiased estimates of the regression coefficients, and they are comparable to those obtained using the known sampling weight benchmark (see Figure 1a). While two approaches are centered near the true values, our proposed method demonstrates modest efficiency gains through reduced variability across most parameters.   When the relationship between $\boldsymbol{X}$ and $\boldsymbol{Z}$ is nonlinear, the integration under homogeneity assumption exhibits pronounced bias and increased variability across multiple regression coefficients, whereas our proposed method yields approximately unbiased estimates that closely align with the know sampling weight benchmark, as illustrated in Figure 1(b). 
\begin{figure*}[htb!]
\centering
\begin{subfigure}[]
  \centering
  \includegraphics[width=\linewidth]{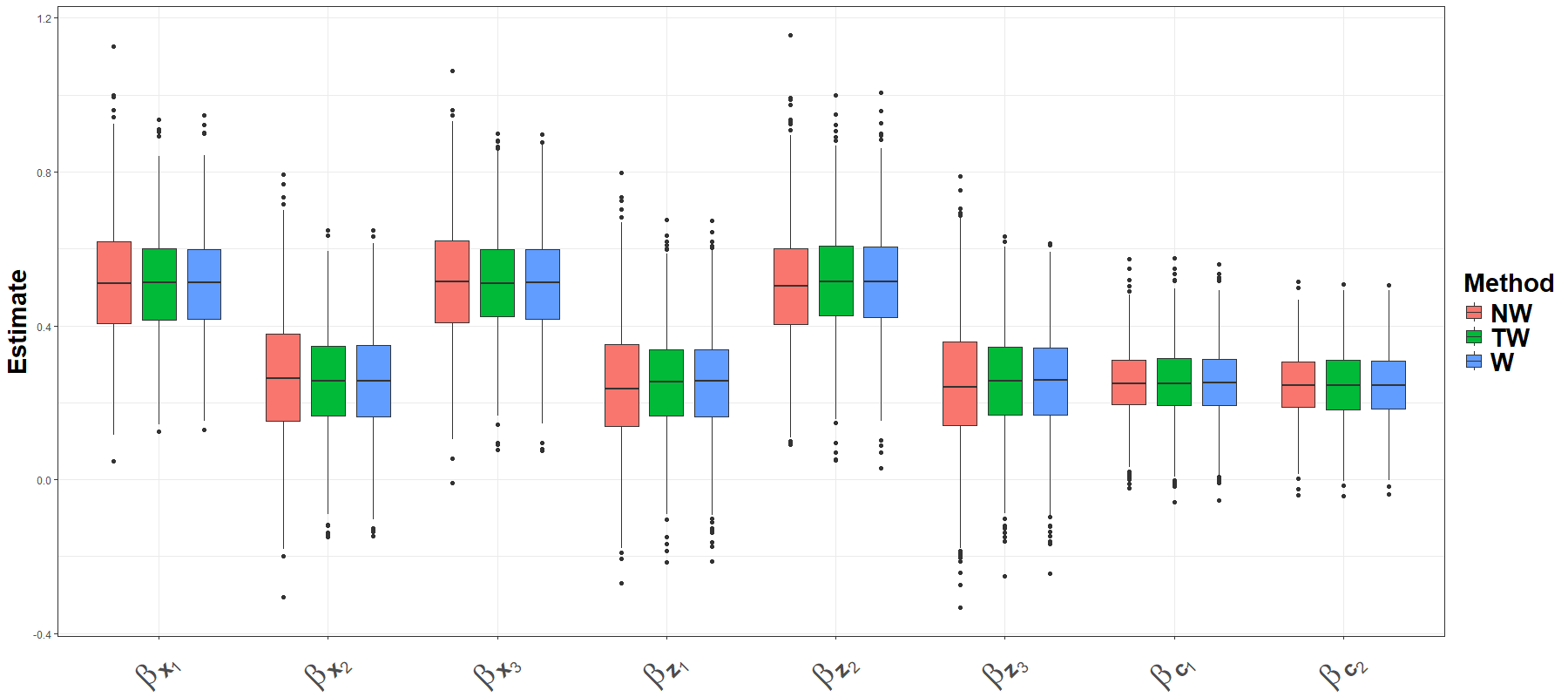}
\end{subfigure}\hfill
 \begin{subfigure}[]
  \centering
  \includegraphics[width=\linewidth]{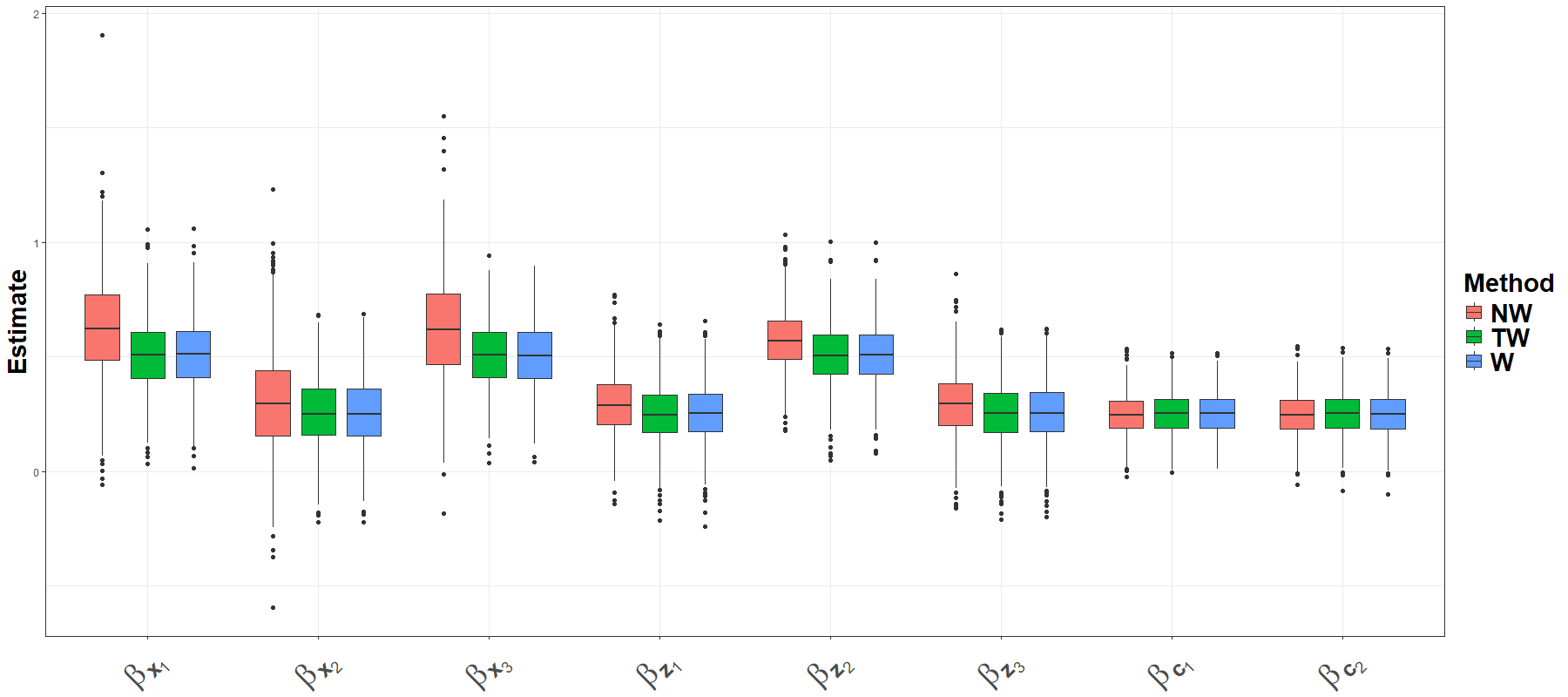}
 \end{subfigure}
\caption{Box-plot of estimates of regression coefficients obtained from three data integration methods under following data generation scenario (a) linear, (b) non-linear relationship between $\boldsymbol{X}$ and $\boldsymbol{Z}$ and assuming sampling bias generated by an additive model. NW, TW, and W indicate methods that integrate summary level data under the homogeneity assumption,
known sampling weights, and the proposed weighting method, respectively.
}
\end{figure*}

 Next, we present the results from the two sets of simulations that evaluate methods under sampling scenario 2 (see Figure 2). Similarly to the previous results, the empirical distributions of estimates indicate that both our proposed approach and integration under homogeneity assumption produce unbiased estimates under a linear relationship (see Figure 2 a). Under a nonlinear relationship, our proposed approach remains approximately unbiased, whereas integration under homogeneity assumption yields biased estimates for most regression coefficients (see Figure 2 (b)).
 
\begin{figure*}[htb!]
\centering
\begin{subfigure}[]
  \centering
  \includegraphics[width=\linewidth]{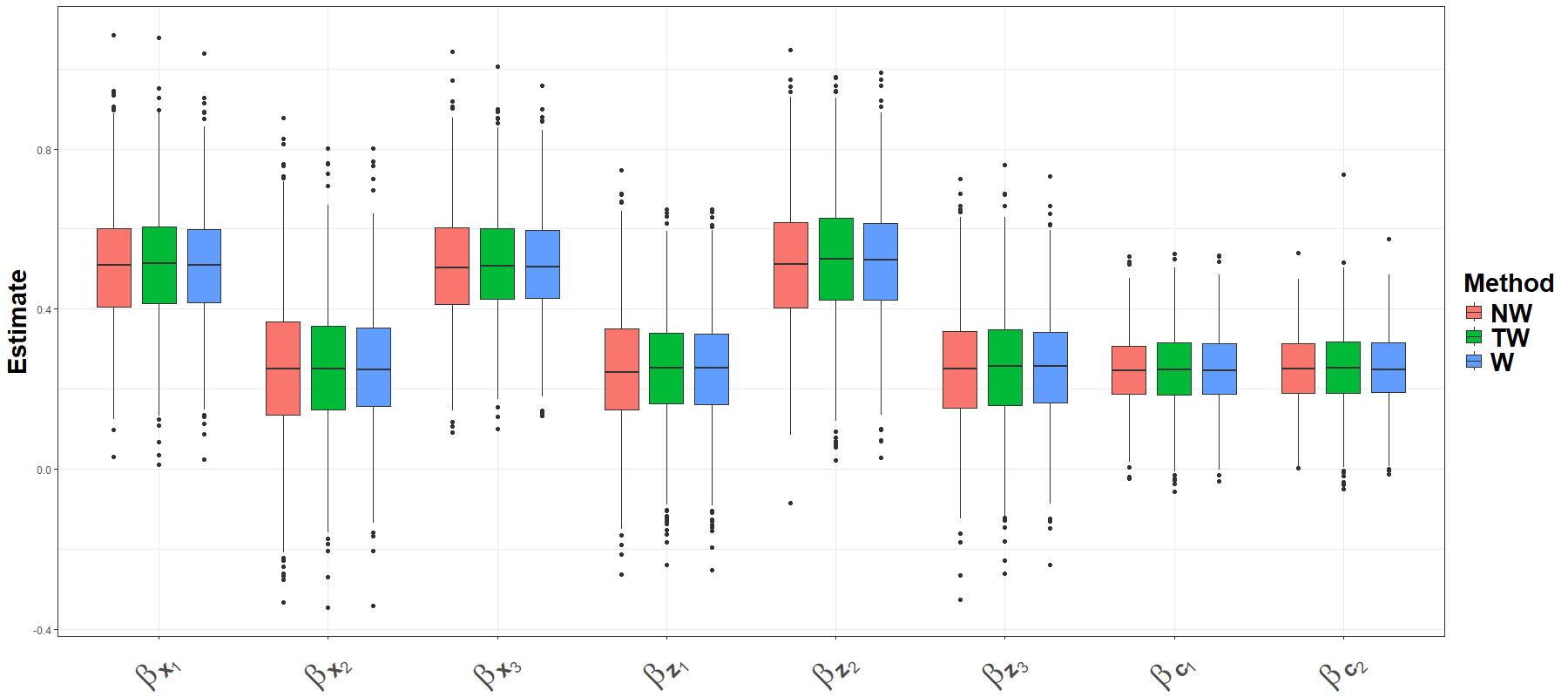}
\end{subfigure}\hfill
 \begin{subfigure}[]
  \centering
  \includegraphics[width=\linewidth]{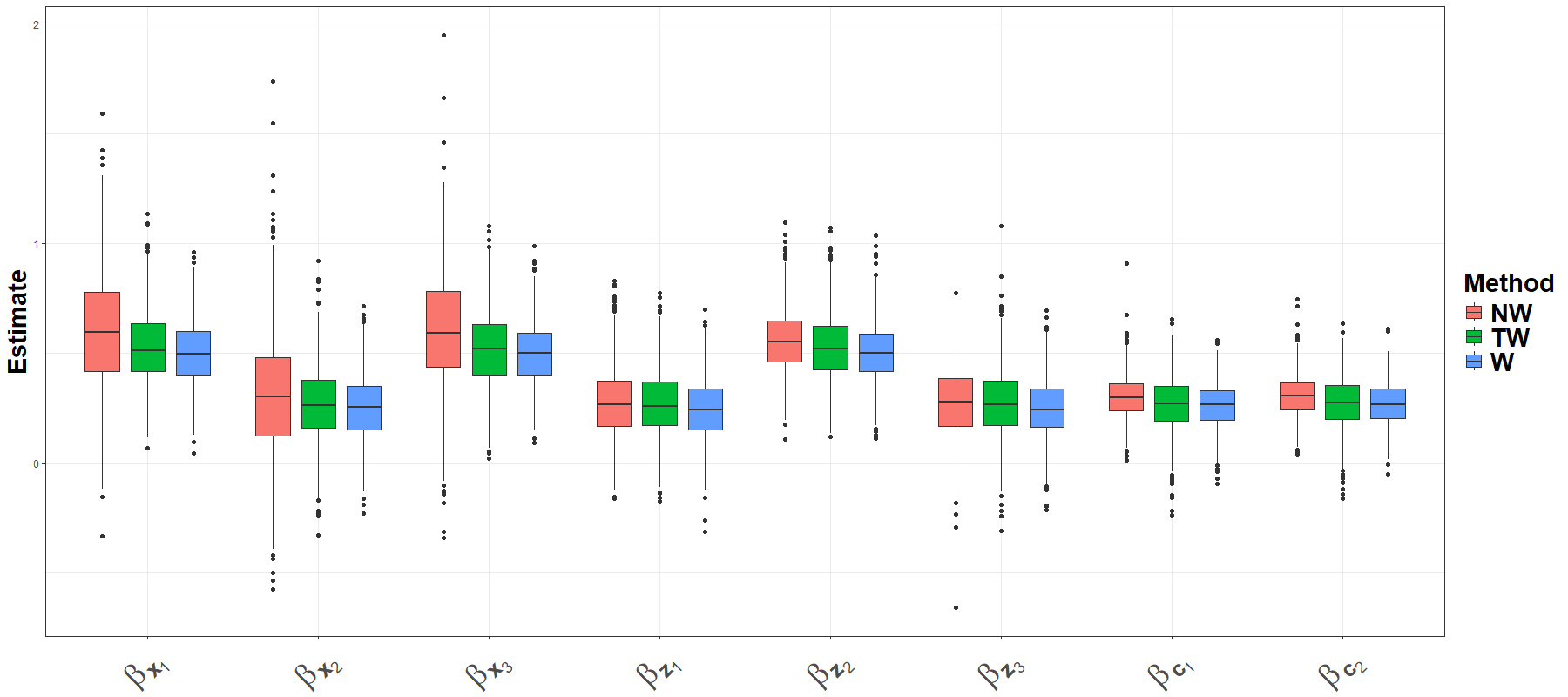}
 \end{subfigure}
\caption{Box-plot of estimates of regression coefficients obtained from three data integration methods under following data generation scenario (a) linear, (b) non-linear relationship between $\boldsymbol{X}$ and $\boldsymbol{Z}$, and assuming sampling bias generated by an additive with interaction model.
NW, TW, and W indicate integration under the homogeneity assumption,
known sampling weights, and the proposed weighting method, respectively
}
\end{figure*}
Lastly, when sampling depends on full data $\bm D$ (i.e. sampling scenario 3), method that integrates data under homogeneity assumption is biased in both linear and non-linear relationships between variables (see Figure 3a and 3b). Our proposed approach shows slight bias under both linear and non-linear associations between covariates, even though candidate calibration weights does not contain correct model.
\begin{figure*}[htb!]
\centering
\begin{subfigure}[]
  \centering
  \includegraphics[width=\linewidth]{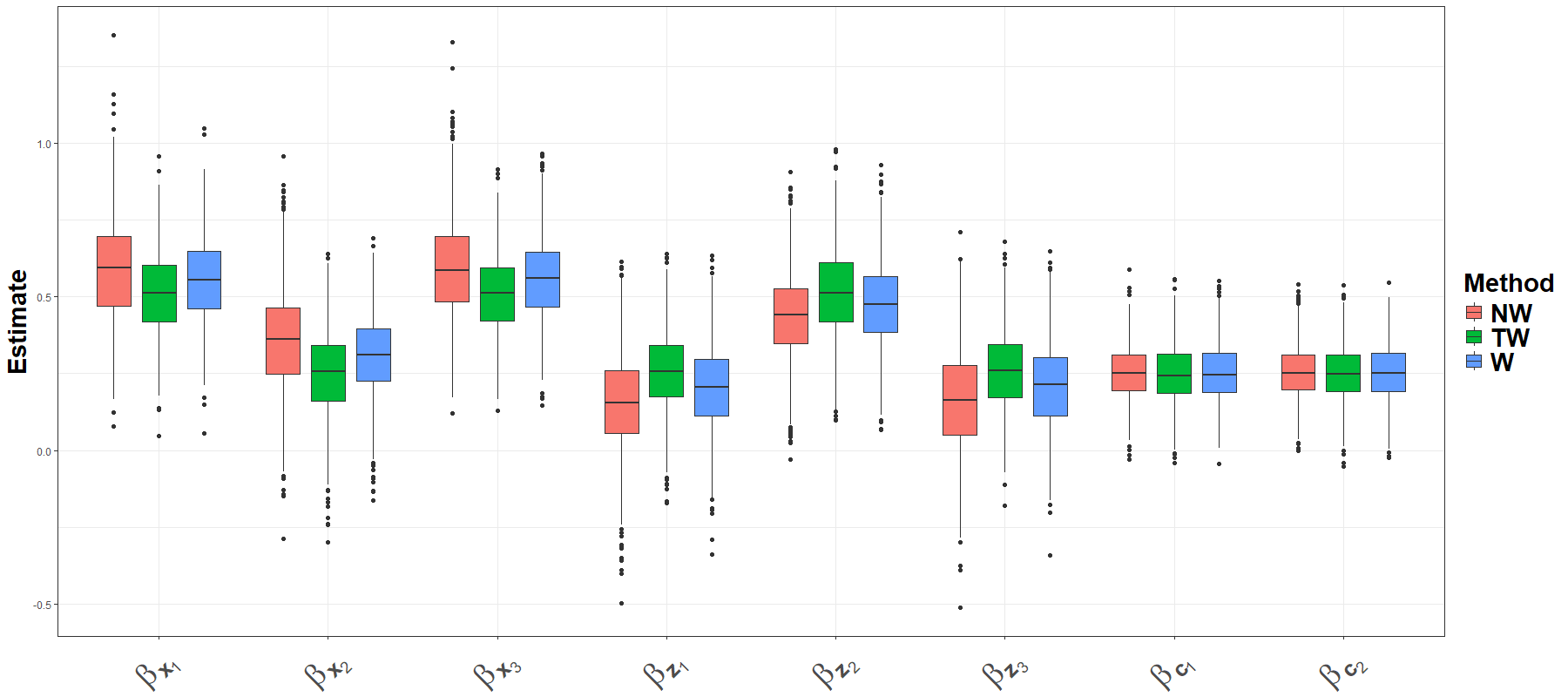}
\end{subfigure}\hfill
 \begin{subfigure}[]
  \centering
  \includegraphics[width=\linewidth]{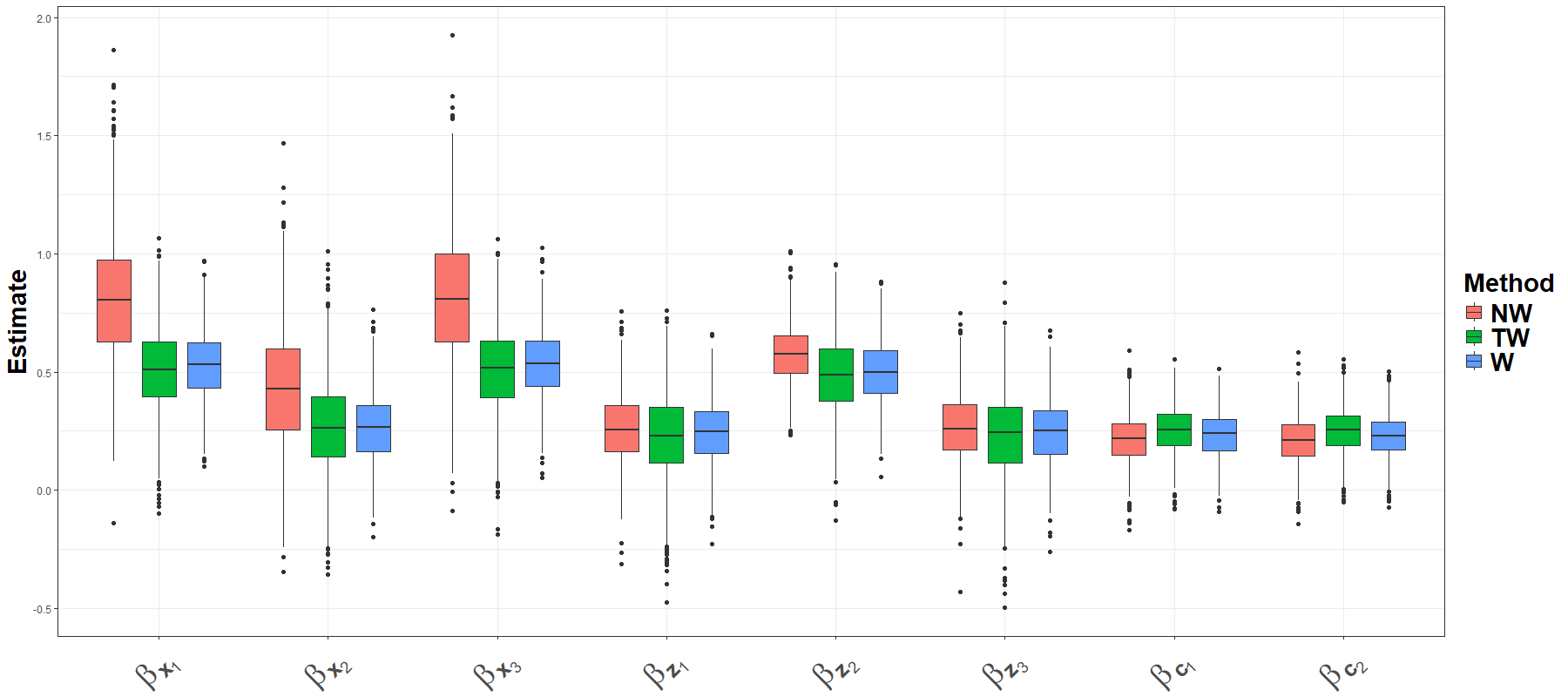}
 \end{subfigure}
\caption{Box-plot of estimates of regression coefficients obtained from three data integration methods under following data generation scenario (a) linear, (b) non-linear relationship between $\boldsymbol{X}$ and $\boldsymbol{Z}$, and assuming sampling bias generated by full model.
NW, TW, and W indicate integration under the homogeneity assumption,
known sampling weights, and the proposed weighting method, respectively.
}
\end{figure*}

We further assess the performance of our proposed weighting method through case–control simulation studies. Figures 4(a) and 4(b) present the estimated regression coefficients under linear and nonlinear relationships between $\boldsymbol{X}$ and $\boldsymbol{Z}$, respectively and assumption of selection bias dependent on additive model. As shown in Figure 4(a), similarly to above results, both methods yield unbiased estimates of the regression coefficients when the relationship between variables is linear. By contrast, in Figure 4(b), the integration under homogeneity assumption produces substantially biased estimates, whereas our proposed approach yields approximately unbiased estimates.

\begin{figure*}[htb!]
\centering
\begin{subfigure}[]
  \centering
  \includegraphics[width=\linewidth]{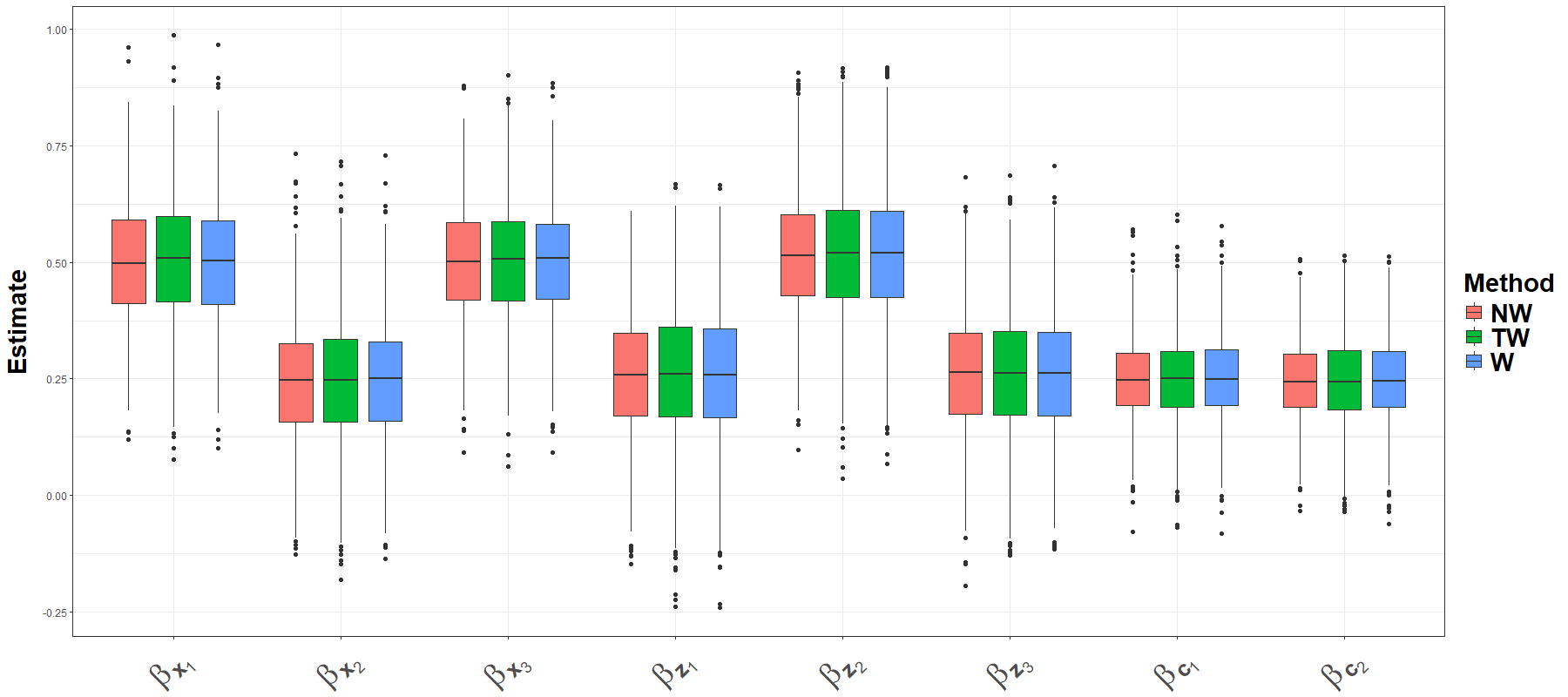}
\end{subfigure}\hfill
 \begin{subfigure}[]
  \centering
  \includegraphics[width=\linewidth]{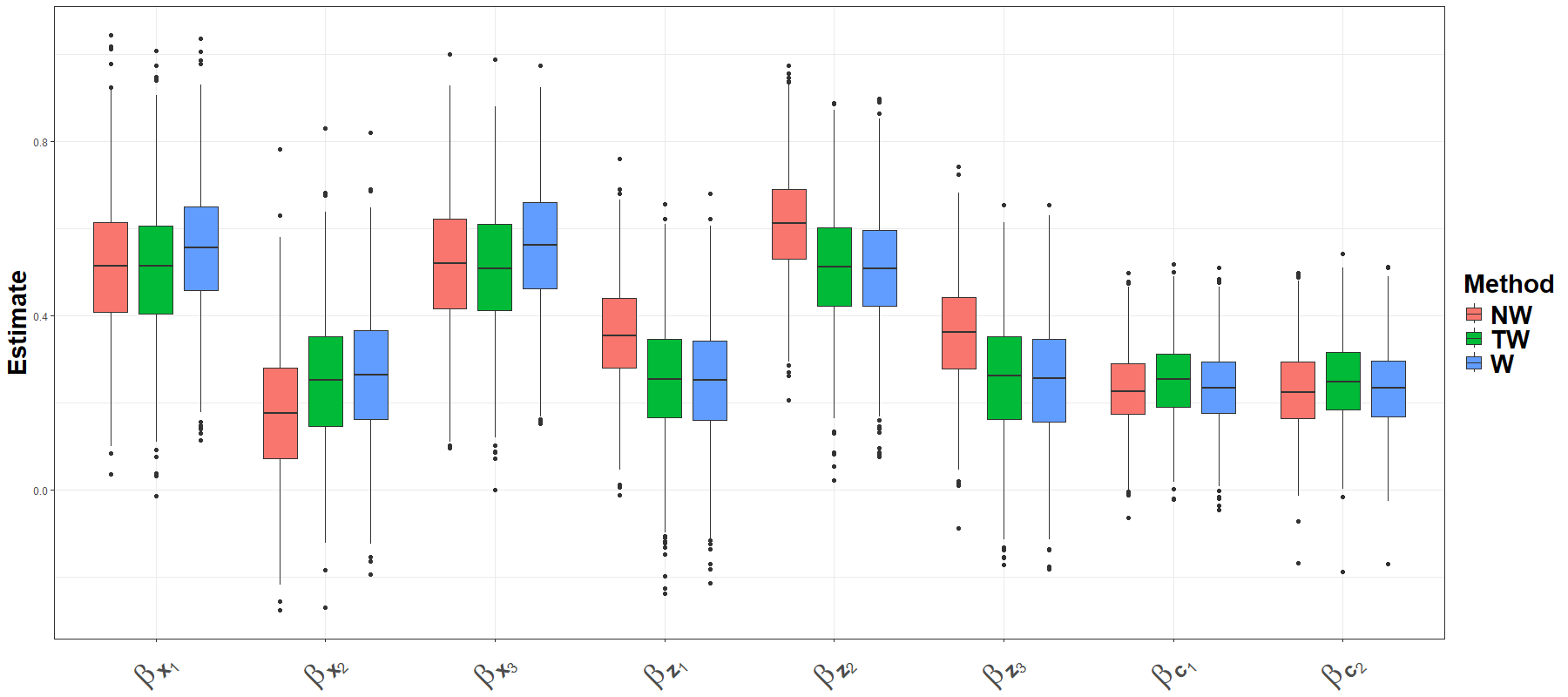}
 \end{subfigure}
\caption{Box-plot of estimates of regression coefficients obtained from three data integration methods under following data generation scenario (a) linear, (b) non-linear relationship between $\boldsymbol{X}$ and $\boldsymbol{Z}$, assuming sampling bias generated by an additive model and case-control sampling to the study 2. NW, TW, and W indicate methods that integrate summary level data under the homogeneity assumption,
known sampling weights, and the proposed weighting method, respectively.
}
\end{figure*}
Lack of significant bias in estimates produced by non-weighting method under additive selection and with interaction models are explained by observation that the correlation between covariates has not been changes significantly. However, when the relationship between variables is non-linear or selection probability depends on all covariates, the correlation between these covariates is changed due to selection procedures. Thus, the integration approach under the homogeneity assumption produces significantly biased estimates.

\clearpage
\section{Motivating example}
 In previous work, \citet{derkach2024mediation} integrated age-specific disease incidence rates, race-specific distributions of three modifiable risk factors, and their associations with disease risk (odds ratios) to estimate adjusted age-specific differences in risks of colorectal cancer incidences between two racial groups. A summary of data sources used in their analysis data is presented in Table 1 with detailed information provided in their article \cite{derkach2024}. 
 
 Here, we briefly outline three sources of the data. The information on the prevalence of modifiable risk factors associated with CRC incidence, including CRC screening history (defined as receipt of sigmoidoscopy and/or colonoscopy within the past 10 years), body mass index (BMI), and regular aspirin use (defined as aspirin intake three or more times per week) for White and Black adults was obtained from the 2018 National Health Interview Survey (NHIS). We obtained sex- and age-specific counts of individuals at risk and newly diagnosed CRC cases in 2018 for non-Hispanic Whites and non-Hispanic Blacks from the NCI's Surveillance, Epidemiology, and End Results (SEER; \url{https://seer.cancer.gov/}) 17 Registries, which cover 26\% of the U.S. population, including 23.6\% of non-Hispanic Whites and 25.5\% of non-Hispanic Blacks.Lastly, study 3 contained estimated effects of modifiable risk factors, including CRC screening history, BMI, and regular aspirin use, on CRC risk over a given time period. These estimates are obtained by computing odds ratios from data on non-Hispanic White men and women drawn from two age-matched case–control studies. Data from the two studies are combined, and conditional logistic regression models accounting for age-group matching are fitted separately for men (1,675 controls and 1,407 cases) and women (1,435 controls and 1,097 cases) \cite{freedman2009colorectal,derkach2024}.

\begin{table}[htbp]
\centering
\caption{Overview of data sources and information used in motivating example.}
\label{tab:data_sources}
\renewcommand{\arraystretch}{1.2}
\begin{tabular}{p{5cm} p{6cm}}
\toprule
\textbf{Data source} & \textbf{Information derived from studies} \\
\midrule
National Health Interview Survey (NHIS) \newline
\url{https://www.cdc.gov/nchs/nhis/nhis_2018_data_release.htm}
&
Individual-level data on modifiable risk factors for both racial groups
\((\boldsymbol{X}, \boldsymbol{Z}, \boldsymbol{C})\), along with survey design variables
(including sampling weights and strata).
\\
\addlinespace[15pt]
Published literature
&
Summary-level data, including age-stratified adjusted effect estimates of modifiable
risk factors \((\boldsymbol{Z}, \boldsymbol{C})\) on colorectal cancer  risk ($Y$),
derived from a matched case--control study. Results are available for White men and women.
\\
\addlinespace[15pt]
NCI Surveillance, Epidemiology, and End Results (SEER; 17 registries) \newline
\url{https://seer.cancer.gov}
&
Summary-level data, including the number of newly diagnosed CRC cases and individuals at risk
in 2018, as well as age-stratified adjusted effect estimates of race \((\boldsymbol{X}, \boldsymbol{C})\) on CRC incidence ($Y$).
\\
\bottomrule
\end{tabular}
\end{table}

\subsection{Main results}
Table 2 presents ORs and 95\% CIs for the association of race, modifiable risk factors with risk of CRC from incomplete data sources (SEER and matched case-control studies), along with ORs estimates from original method that assumes homogeneity between three studies (\cite{derkach2024mediation}) and re-estimated ORs obtained using the proposed weighting method. 

As reported by \citet{derkach2024mediation}, under the homogeneity assumption between studies, adjustment for modifiable risk factors (CRC screening, BMI, and aspirin use) resulted in only a negligible attenuation of the association between race and CRC risk, indicating that these factors explain little of the observed racial disparity. The one explanation there is potential bias in the estimates due to potential heterogeneity between data sources. Specifically, we apply our new method while adjusting for potential heterogeneity between NHIS and SEER databases with the plans of re-analyzing data when additional auxiliary information will be provided for the matched case-control study. We estimated parameters for calibration weights from covariance matrix of the regression coefficient estimates obtained on the SEER database using 80\%/20\% train and test framework from two candidate models: additive and full model (see Section 3.2). The full and additive model fitted well with difference between observed and predicted Fisher information matrix being very close. The full model had slightly lower value, and was selected in the final analysis. The fourth column of the Table 2 shows results of our approach, which is very close to one under homogeneity. 

\begin{table}[H]
			\centering
            \captionsetup{width=\linewidth}
            \caption{Estimate effect of race ($\boldsymbol{X}$), modifiable risk factors ($\boldsymbol{Z}$), and age ($\boldsymbol{C}$) on CRC risk in men from two data sources and full joint models using the Derkach et al. and proposed methods.}
\label{tab:data_sources}
            \large
            \renewcommand{\arraystretch}{1.25}
			\resizebox{\linewidth}{!}{	
					\begin{tabular} {p{0.48\linewidth}|p{0.25\linewidth}|p{0.25\linewidth}|p{0.27\linewidth} |p{0.27\linewidth}}
						\hline
						Variables &	ORs estimated from SEER (2018) &	{Estimates of ORs provided by published by external study}$^*$ &
                        \textbf{Estimates of ORs for the full model(Derkach et al.)} &
					\textbf{Estimates of ORs for the full model(proposed approach)} 
						\\
						\hline
						Colonoscopy/sigmoidoscopy within last ten years	& $-$	&0.68 (0.63,0.74)	& $~~~~~$0.68 (0.63,0.74) & $~~~~~$0.68 (0.63,0.74)\\
						BMI
                        
                        $~~$(Ref: Healthy ($\leq24.9$ kg/m2)) &&& \\

						$~~$Overweight ($25-30$ kg/m2) &	$-$	&0.98 (0.88,1.08)	& $~~~~~$0.98 (0.88,1.08) & $~~~~~$0.98 (0.88,1.08)\\
						$~~$Obese ($>30$ kg/m2)  &	$-$	&1.3 (1.16,1.47)	& $~~~~~$1.3 (1.16,1.47) & $~~~~~$1.29 (1.16,1.47)\\
						Regular aspirin use &	$-$	& 0.89 (0.82,0.96)&$~~~~~$0.88 (0.82,0.96) &$~~~~~$0.88 (0.82,0.96) \\
						\hline
						NH White vs NH Black &	0.77 (0.73,0.81) &	$NA$	& $~~~~~$0.79 (0.73,0.81) & $~~~~~$0.78 (0.73,0.81)\\
						Age (Ref: 70-74) &&& \\
						$~~$40-49 &	0.14 (0.12,0.17)&	Not-reported	& $~~~~~$0.11 (0.12,0.17) & $~~~~~$0.11 (0.12,0.17)\\
						$~~$50-59 &	0.42 (0.40,0.44)&	Not-reported	& $~~~~~$0.37 (0.40,0.44) & $~~~~~$0.37 (0.40,0.44) \\
						$~~$60-69 &	0.64 (0.62,0.68)&	Not-reported& $~~~~~$0.62 (0.62,0.68) & $~~~~~$0.62 (0.62,0.68)\\
						Race*Age of 40-50 & 	1.36  (1.16,1.59) &	Not-reported & $~~~~~$1.33 (1.16,1.59) & $~~~~~$1.33 (1.16,1.59)\\
						\hline
				\end{tabular}}
		\end{table}

\section{Discussion}
Standard regression models break down when data from multiple sources are integrated such that no subject has fully observed data. Recent work has proposed several methods for combining individual-level and summary-level data to improve the efficiency of regression parameter estimation under a homogeneity assumption across data sources \cite{chatterjee2016, li2020, vo2019novel}; however, this assumption is rarely satisfied in practice. We have proposed a weighting approach for integrating summary-level with individual-level data from heterogeneous data sources to estimate regression coefficients in settings that commonly arise when covariates and outcomes are not jointly observed within a single study.

We demonstrated that all regression parameters are identifiable under the generalized linear models (2.1–2.3), and that the resulting estimators converge in probability to the true parameter values. The proposed method has several properties. Importantly, simulation studies across multiple biased sampling scenarios (Scenarios 1–3) indicate that the proposed approach yields unbiased estimates and confidence intervals with nominal coverage. Moreover, under linear covariates relationships, the proposed approach attains efficiency comparable to the previous method \cite{derkach2024integrating}; however, under nonlinear relationships, only the proposed method yields unbiased estimates. In simulation studies for case–control settings, the proposed method yields unbiased estimates of the regression parameters under both linear and nonlinear relationships between covariates.

Our approach provides an exploratory framework for quantifying the contribution of established risk factors to racial disparities in CRC incidence in the United States. Our findings suggest that CRC screening history within the past 10 years, BMI, and regular aspirin use accounted for only a small proportion of the racial disparities observed in 2018. Although the proposed and existing methods yield similar results—likely due to the use of population-based survey data—the use of SEER cancer registries and NHIS survey ensures that the findings are broadly generalizable to the U.S. population.

Our proposed framework has several limitations. First, the model specified in sections 2.3 assume no interaction between $(\boldsymbol{X}, \boldsymbol{Z}, \boldsymbol{C})$. In the absence of additional data containing all three measurements, inclusion of an unknown interaction parameter in the outcome model of Section 2.3 leads to identifiability issues. Second, we have assumed that the conditional distribution $Y$ given $\bm D$ is the same across data sources and $Y$ belongs to exponential family. For alternative parametric model specifications, the proposed weighting approach is still applicable for estimating the regression parameters as it reverse covariate shifts between studies. Third, extreme weights may occur when there is limited overlap between the source of summary-level data and individual-level dataset or sample size is low. Although alternative approaches have been proposed to address this issue, they can be highly sensitive to assumptions imposed on outcome model–based effect modification  \cite{vo2023cautionary}. Finally, although our focus is on generalized linear models, we plan to extend the approach to accommodate other classes of parametric models, including survival models. These extensions will be pursued in future work.Despite these limitations, the proposed method provides a practically important and novel tool for integrating information from heterogeneous data sources with partially observed data.

\clearpage
\bibliographystyle{chicago}
\bibliography{references}

\end{document}